%% file: paper.tex
\documentclass[sigconf]{acmart}

\usepackage{url}
\usepackage[utf8]{inputenc}
\usepackage{xcolor}
\usepackage[normalem]{ulem}

\title{Peak Forecasting for Battery-based Energy Optimizations in Campus Microgrids} 

\acmYear{2020}\copyrightyear{2020}
\setcopyright{acmlicensed}
\acmConference[e-Energy '20]{The Eleventh ACM International Conference on Future Energy Systems}{June 22--26, 2020}{Virtual Event, Australia}
\acmBooktitle{The Eleventh ACM International Conference on Future Energy Systems (e-Energy '20), June 22--26, 2020, Virtual Event, Australia}
\acmPrice{15.00}
\acmDOI{10.1145/3396851.3397751}
\acmISBN{978-1-4503-8009-6/20/06}


\author{Akhil Soman}
\affiliation{ \institution{University of Massachusetts, Amherst}}
\email{asoman@umass.edu}

\author{Amee Trivedi}
\affiliation{ \institution{University of Massachusetts, Amherst}}
\email{amee@cs.umass.edu}

\author{David Irwin}
\affiliation{ \institution{University of Massachusetts, Amherst}}
\email{irwin@ecs.umass.edu}

\author{Beka Kosanovic}
\affiliation{ \institution{University of Massachusetts, Amherst}}
\email{kosanovic@umass.edu}

\author{Benjamin McDaniel}
\affiliation{ \institution{University of Massachusetts, Amherst}}
\email{bmcdaniel@umass.edu}

\author{Prashant Shenoy}
\affiliation{ \institution{University of Massachusetts, Amherst}}
\email{shenoy@cs.umass.edu}

\begin{document}

\renewcommand{\shortauthors}{Soman et al.}
\input{abstract}

\maketitle

\input{introduction}
\input{background}
\input{model}

\input{results}
\input{conclusions.tex}

\bibliographystyle{acm}
\bibliography{paper}

\end{document}

%% file: abstract.tex
\begin{abstract}
Battery-based energy storage has emerged as an enabling technology for a variety of grid energy optimizations, such as peak shaving and cost arbitrage. A key component of battery-driven peak shaving optimizations is peak forecasting, which predicts the hours of the day that see the greatest demand. While there has been significant prior work on load forecasting, we argue that the problem of predicting periods where the demand peaks for individual consumers or micro-grids is more challenging than forecasting load at a grid scale. We propose a new model for peak forecasting, based on deep learning, that predicts the $k$ hours of each day with the highest and lowest demand. We evaluate our approach using a two year trace from a real micro-grid of 156 buildings and show that it outperforms the state of the art load forecasting techniques adapted for peak predictions by 11-32\%. When used for battery-based peak shaving, our model yields annual savings of \$496,320 for a 4 MWhr battery for this micro-grid. 
\end{abstract}

%% file: introduction.tex
\section{Introduction}
\label{sec:introduction}

Energy storage has emerged as a key enabling technology for various grid and energy optimizations such as peak load shaving and energy cost arbitrage. Energy storage is also becoming popular for smoothing energy generation from intermittent sources such as solar and wind. The cost of battery-based storage has been declining steadily over the years, and the penetration of battery storage, while still nascent, is poised to grow sharply in the coming years. 

Battery-based storage is particularly attractive for large energy consumers such as commercial and industrial customers or micro-grids. Unlike the majority of residential consumers that pay a flat rate for electricity, such customers pay demand charges, which is effectively a surcharge based on their peak usage. The use of batteries to flatten the energy consumption during peak demand hours can be an effective mechanism for reducing these demand charges. For example, our campus micro-grid has recently deployed a 4 MWhr battery for the sole purpose of peak load reduction and the incurred demand charges.

An essential ingredient of any peak load shaving technique is the prediction of when the peak demand will occur each day so that the battery can be operated during those hours to reduce the peak power draw from the grid.  We refer to this problem as {\em peak forecasting}, which is related to, but distinct from, the problem of load forecasting. Load forecasting is a well studied problem in the literature \cite{kim2019short,ryu2017deep,khan2016load,ghasemi2016novel,hippert2001neural,zheng2017electric,laouafi2016daily} and involves predicting a time series of future demand using past history and parameters such as weather. In contrast, peak forecasting is concerned with predicting specific hours of the day when the demand will peak. We note that any load forecasting method can be trivially modified for peak forecasting by first predicting a time series of the demand and then sorting the demand to determine the top-k peak hours over the prediction window. However, we argue that load forecasting methods were designed for grid-level predictions where the demand varies smoothly. Peak forecasting is applied to individual consumers, or micro-grids, where the daily demand exhibits higher variations. Conventional load forecasting may be less sensitive to peaks in demand, while a peak forecasting method that is solely concerned with determining peak demand periods, rather than detailed predictions of a time-series of demand, maybe more effective.

Motivated by these observations, in this paper, we present a new peak forecasting technique that is designed for predicting the top-k and bottom-k high and low demand hours for each day. Our prediction model is based on a deep learning-based Long Short Term Memory (LSTM) approach that is tailored for micro-grids or large commercial customers that exhibit higher stochasticity in their demand than grid-scale demand variations. Such a model can be directly used for battery control and peak shaving by operating the battery during the predicted top-k hours and charging the battery during the bottom-k off-peak hours. In designing our approach, we make the following contributions.

First, we formulate the problem of peak forecasting and present an LSTM model tailored for predicting the $k$ high and low demand periods during each day for any configurable $k$. Second, we compare our approach with state of the art load forecasting approaches, suitably adapted for peak forecasting, using a two-year-long demand trace from our campus microgrid comprising 150 buildings. Our results show that our approach can outperform the state of the art methods by 11-32\%. We conduct a case study of a campus micro-grid comprising a 4 MWhr battery where we apply our model for battery control to perform peak shaving and show annual energy savings of \$496,320. 
Finally, we implement our approach on a Raspberry Pi to demonstrate its feasibility of running in embedded battery controllers for autonomous operations and also provide an open-source implementation of our model as a library.




%% file: background.tex
\section{Background}


Our work focuses on larger energy consumers such as office or university campuses, shopping malls, convention centers, or manufacturing facilities. We assume that such customers are subjected to demand charges, also known as peak surcharges, that is a surcharge paid on the monthly electricity bill based on the peak usage of that customer. 
As a result, the customer is financially motivated to flatten the peak usage as much as possible---the smoother the demand, the lower the peak charge.  With the emergence of battery-based storage technologies, it has become feasible to reduce the grid-observed peak demand without actually changing the underlying consumption patterns. This is done by operating the battery during the peak demand to absorb a portion, or all, of the peak. Our work assumes that customers who wish to employ such optimizations have deployed battery-based  storage of a certain known capacity. 

Past work on using batteries for grid energy optimizations falls broadly into two categories: energy arbitrate and peak shaving.  When customers are subjected to the time of use (TOU) pricing, with different prices during pre-defined peak and off-peak periods, batteries enable energy to arbitrate where the battery charges during cheaper off-peak hours and discharges during peak hours to reduce bills. Past work has cast this problem as an optimization problem where load forecasting is used to estimate future demand and the optimization determines the optimal amount of charging or discharging to maximize savings.  Peak shaving \cite{barker2012smartcap,reihani2016load,rahimi2013simple, shi2017using,barth2019shaving} is a different type of energy optimization that is designed to address peak demand charges---in this case, the customer needs to predict when their demand is likely to peak, and operate the battery during this period to ``clip'' the peak. In this case, it is more critical to determine when the local demand from the customer will peak during each day, a problem we refer to as peak forecasting.  As noted earlier, the problem of peak forecasting is a related but distinct problem from load forecasting---in the former, we need to predict the top-k hours when demand will be the greatest, while in the latter, we need to predict a time-series of estimated demand. 


Load forecasting is a well-studied problem with many decades of research.  Past work in the area falls into two broad categories: use of time series forecasting methods \cite{hagan1987time, sadaei2017short, qiu2017empirical} and, more recently, use of neural nets and deep learning methods \cite{qi2019energyboost, ghiassi2016joint, ryu2017deep}. Regardless of the method, all load forecasting techniques use past history and parameters such as weather to estimate future demand. Load forecasting methods are known to be very accurate for predicting grid-scale demand where the variations are ''smooth'' but have higher errors when predicting demand for individual consumers where demand has higher stochasticity. 

Peak forecasting \cite{dowling2018coincident, chandio2019gridpeaks} is less well studied than load forecasting. The problem was studied in \cite{kazhamiaka2016practical} with the goal of predicting the top-5 peak days in each year for the region of Ontario \cite{jiang2016analyzing}; we seek to perform peak forecasting at the shorter time-scale of hours, which is more challenging since hourly individual demand has higher variations than aggregated daily grid demand. As mentioned earlier, one baseline approach is to take any load forecasting approach and trivially modify it for peak forecasting by sorting the predicted time series and choosing the top-k hours. As we will show experimentally, such an approach can yield higher errors. 

%% file: model.tex
\section{Peak Forecasting Model}
\label{sec:model}

In this section, we present our 
model for peak forecasting.


\subsection{Peak Forecasting using LSTM}
\begin{figure}[t]
\includegraphics[height=1.8in]{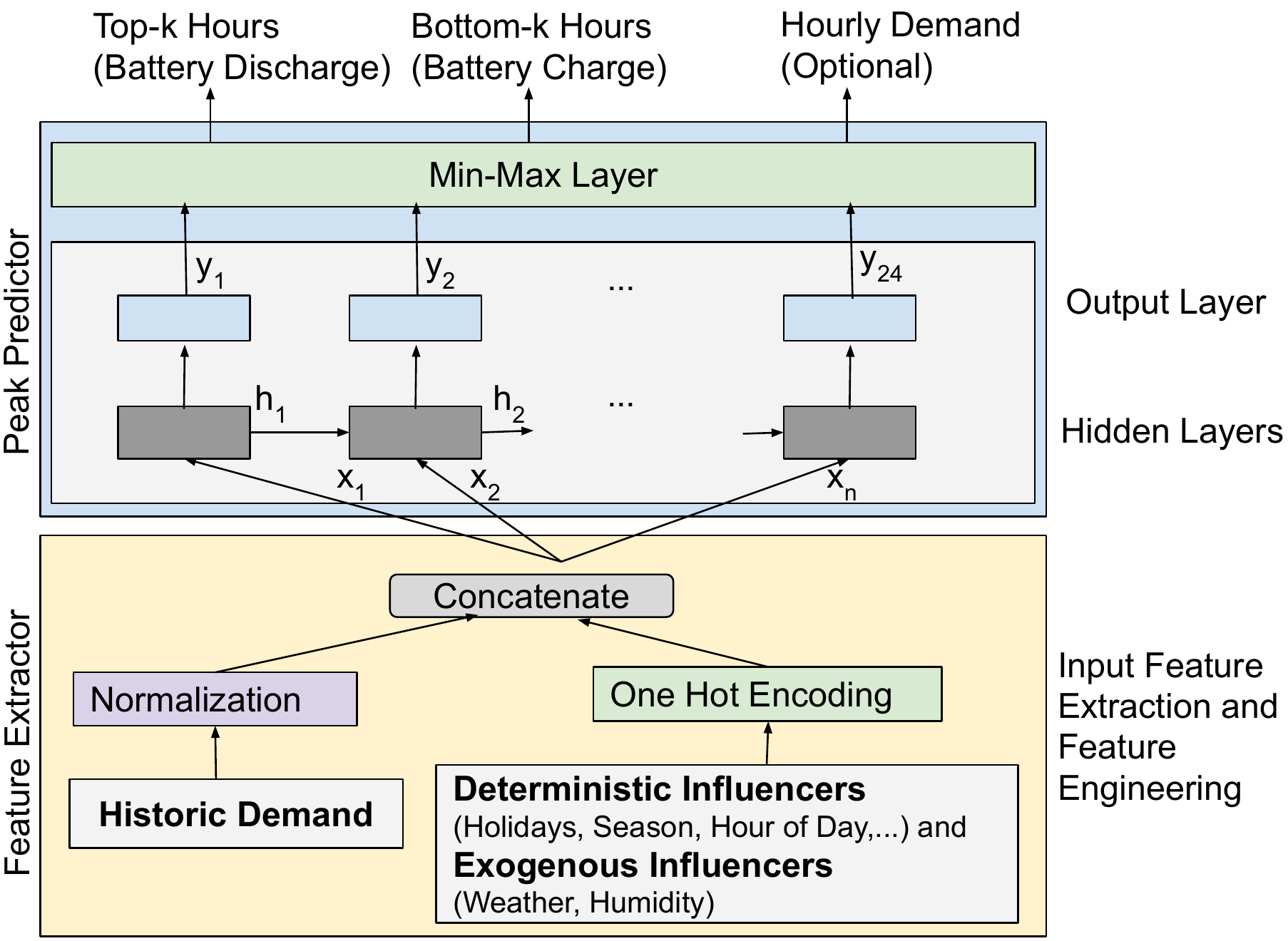}
\vspace{-0.1cm}
\caption{LSTM based demand predictor}
\label{fig:lstm}
\vspace{-0.3cm}
\end{figure}

The main objective of the peak forecasting model is to predict the top-k and bottom-k hours of daily demand. 
Figure \ref{fig:lstm} shows the architecture diagram of our model. Our model has 2 main modules the \emph{Feature Extractor} and \emph{Peak Predictor}.

\textbf{Feature Extractor :} We use the historic demand as the input to the model along with few engineered features to improve the accuracy of the model. We add deterministic influencers such as holidays, the hour of the day, season type (Fall, Winter, Spring, Summer), holidays and exogenous influencers such as weather and humidity to improve the accuracy of our model. 
 We encode all influencer features using One Hot Encoding, a method of converting categorical variables to vector form and normalize the historic hourly demand. Then, the one hot encoded features and normalized historic demand are concatenated and fed as input to the peak predictor.



\textbf{Peak Predictor :} The peak predictor is a stack of 4-layer Long Short Term Memory(LSTM), which is a variant of a Recurring Neural Network (RNN) and a Min-Max Layer. The objective of the peak predictor is to predict the top-k and bottom-k hours over the next 24 hours given the historic hourly demand for the past 2 days along with other engineered features. To predict the hourly demand we use a variant of a Recurring Neural Network (RNN) called LSTM (Long Short Term Memory) since it is well known that an LSTM \cite{hochreiter1997long, hochreiter1998vanishing} shows superior performance in learning sequential and long term dependencies in data.  



As shown in \cite{hernandez2012study}, grid demand displays a high degree of temporal correlations along with sequential dependency. RNNs have been shown to capture the sequential dependencies very well. However, as the sequence length increases, long term dependencies can be lost due to the vanishing gradient problem \cite{hochreiter1998vanishing}. To overcome this, we follow previous work and use the Long Short Term Memory (LSTM) variant of RNN \cite{hochreiter1997long} in our LSTM based demand prediction model. The LSTM incorporates three additional matrices inside the recurrence which act as a gating mechanism, selectively allowing information flow from previous timesteps as a function of the current timestep. The gates $f_t$, $i_t$ and $o_t$ are defined as functions of the input $x_t$ and previous hidden representation $h_{t-1}$ as follows.
\begin{align}
f_t &= \sigma_g(W_fx_t + U_fh_{t-1} + b_f) \\
i_t &= \sigma_g(W_ix_t + U_ih_{t-1} + b_i) \\
o_t &= \sigma_g(W_ox_t + U_oh_{t-1} + b_o)
\end{align}
Here, matrices $W$ and $U$ and bias vectors $b$ are all model parameters learned through supervised training. $\sigma_g$ denotes the sigmoid activation function, which normalizes the outputs to have values in [0, 1]. These representations are combined to produce the hidden representation $h_t$ as follows:
\begin{align}
c_t &= f_t\circ c_{t-1} + i_t \circ \sigma_c(W_cx_t + U_ch_{t-1} + b_c) \\
h_t &= o_t \circ \sigma_h(c_t)
\end{align}
where $\circ$ denotes element-wise vector multiplication. Activations $\sigma_c$ and $\sigma_h$ use the tanh function to normalize outputs into the range [-1, 1], following previous work.

The final layer of our model is a min-max layer that labels each hour as T (if it is a top-k hour), B (if it is a bottom-k hour), or N (for neither). As shown in Figure \ref{fig:lstm}, our models can optionally output the predicted hourly demand (load forecast) in addition to the top-k and bottom-k hour labels.


\subsection{Applying the Model for Battery Control}
Our peak forecasting model  can then be used for battery-based peak shaving. For example, the model output can be used to directly control the battery, where the battery discharges during the top-k hours and charges back to full in the bottom-k hours. Model predictions can also be incorporated into more sophisticated battery control algorithms that incorporate solar renewables, battery lifetime, and other factors for energy optimizations.

%% file: results.tex
\section{Evaluation}
\label{sec:evaluation}
In this section, we experimentally evaluate our model using a real 2-year demand trace from a campus micro-grid of 156 buildings. The trace sees a temperature range of -9.5F to 97F and demand variations between 9,934 kW to 26,219 kW. 
For our evaluation, we use two LSTM models, a 2 layer model and a 4 layer model that we implement using Keras with TensorFlow ~\cite{abadi2015tensorflow} backend. The 2 layer NN has 100 and 80 neurons, while 4 layer NN has 100, 90, 80, and 70 neurons in the hidden layers ordered from lower to the upper layer. A grid search was performed for the selection of hyper-parameters. For training, we use Adam optimizer with an adaptive learning rate of 0.1 to 0.005 with 0.2 drop out. 

For the purpose of comparison, we use 4 load forecasting models that we adapt for peak forecasting by sorting their outputs and choosing the top and bottom $k$ values.  We use a linear regression model as baseline and two state-of-the-art load forecasting models Custom ARIMA \cite{kim2019short}, and Artificial Neural Network (ANN) model.  We tune the ANN hyper-parameters using grid-search and train all models using the same campus microgrid dataset.

\begin{table}[ht]
\begin{minipage}[l]{0.36\linewidth}
    \centering
\begin{small}
   \begin{tabular}{lc} \toprule
        Model &  MAPE \\ \midrule
        LSTM-2 Layer & 4.1 \\ 
        LSTM-4 Layer &  3.7\\
        Linear Regression & 5.9\\
        ANN &  3.8\\
        ARIMA & 5.0 \\ \bottomrule
    \end{tabular}
  \end{small} 
    \caption{Demand Prediction Accuracy}
    \label{tab:demand_prediction}
    \vspace{-0.6cm}
\end{minipage}
\hfill
\begin{minipage}[c]{0.54\linewidth}
\centering
\includegraphics[width=1.6in]{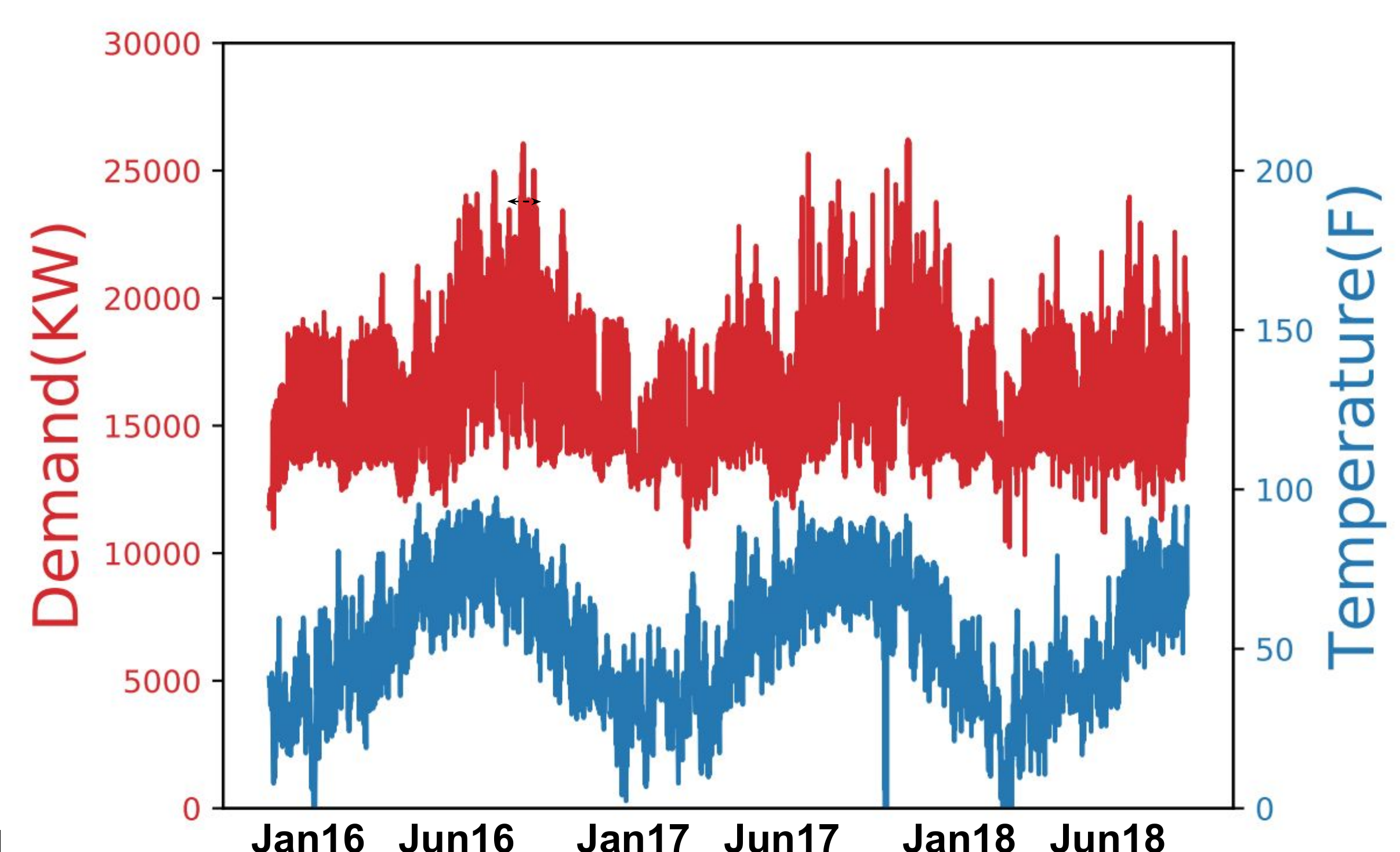}
\captionof{figure}{Dataset temperature range and variance}
\vspace{-0.6cm}
\label{fig:Dataset_Temperature}
\end{minipage}%
\end{table}

\subsection{Baseline Evaluation}
Although our focus is on peak forecasting, we first compare the Mean Absolute Percentage Error (MAPE) of all approaches for predicting the demand time series. As shown in Table~\ref{tab:demand_prediction} we find that our 4-layer LSTM model has the least MAPE and outperforms even state of the art load forecasting approaches.  The ANN approach, with a MAPE of 3.8, is a close second, while linear regression has the highest error. Minor improvements in MAPE score has a significant impact on peak prediction, which results in substantial cost savings.

\begin{figure}[h]
    { \setlength{\tabcolsep}{2pt}
    \begin{tabular}{cc}
    \includegraphics[width=1.5in]{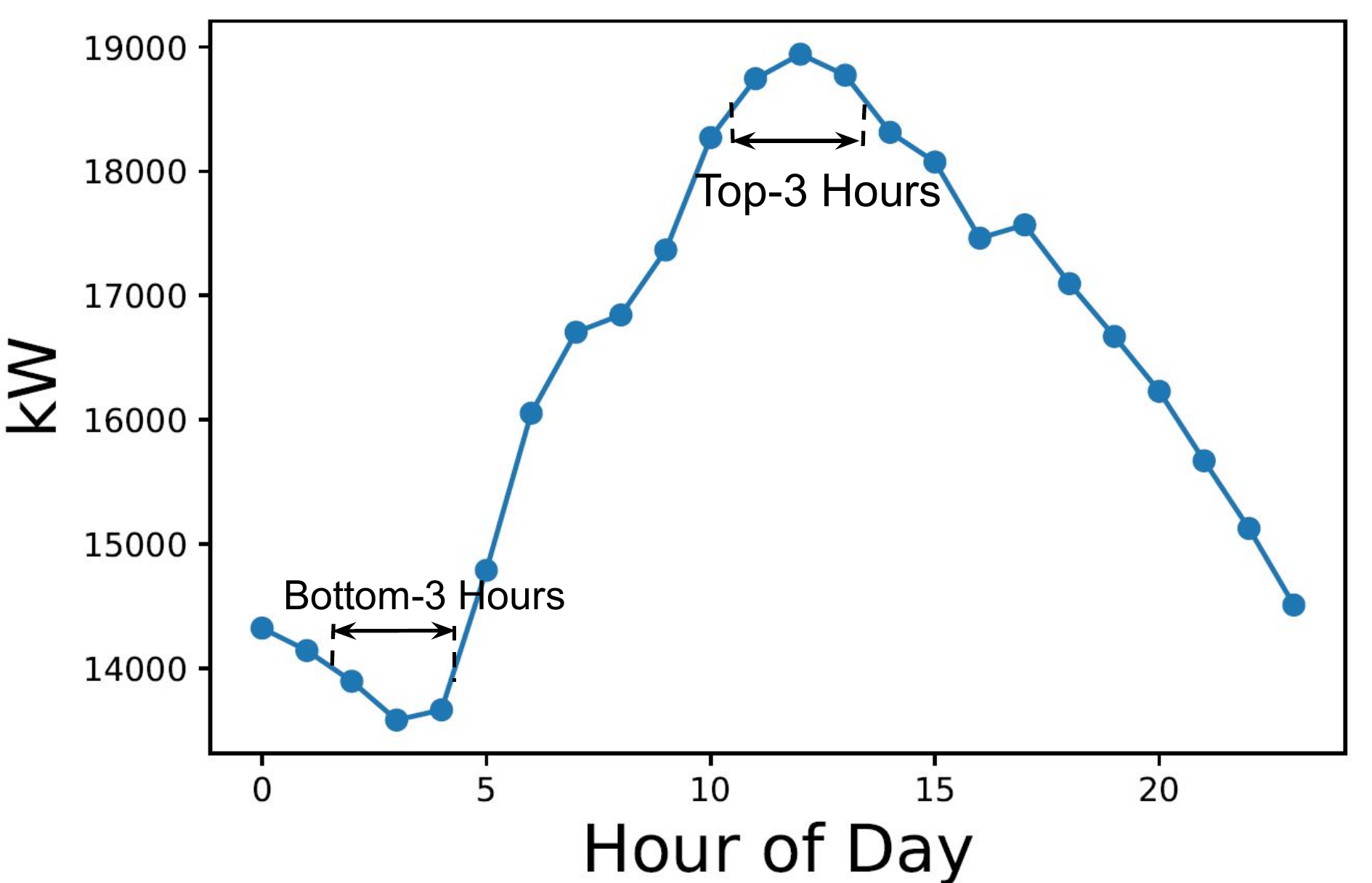} &
    \includegraphics[width=1.5in]{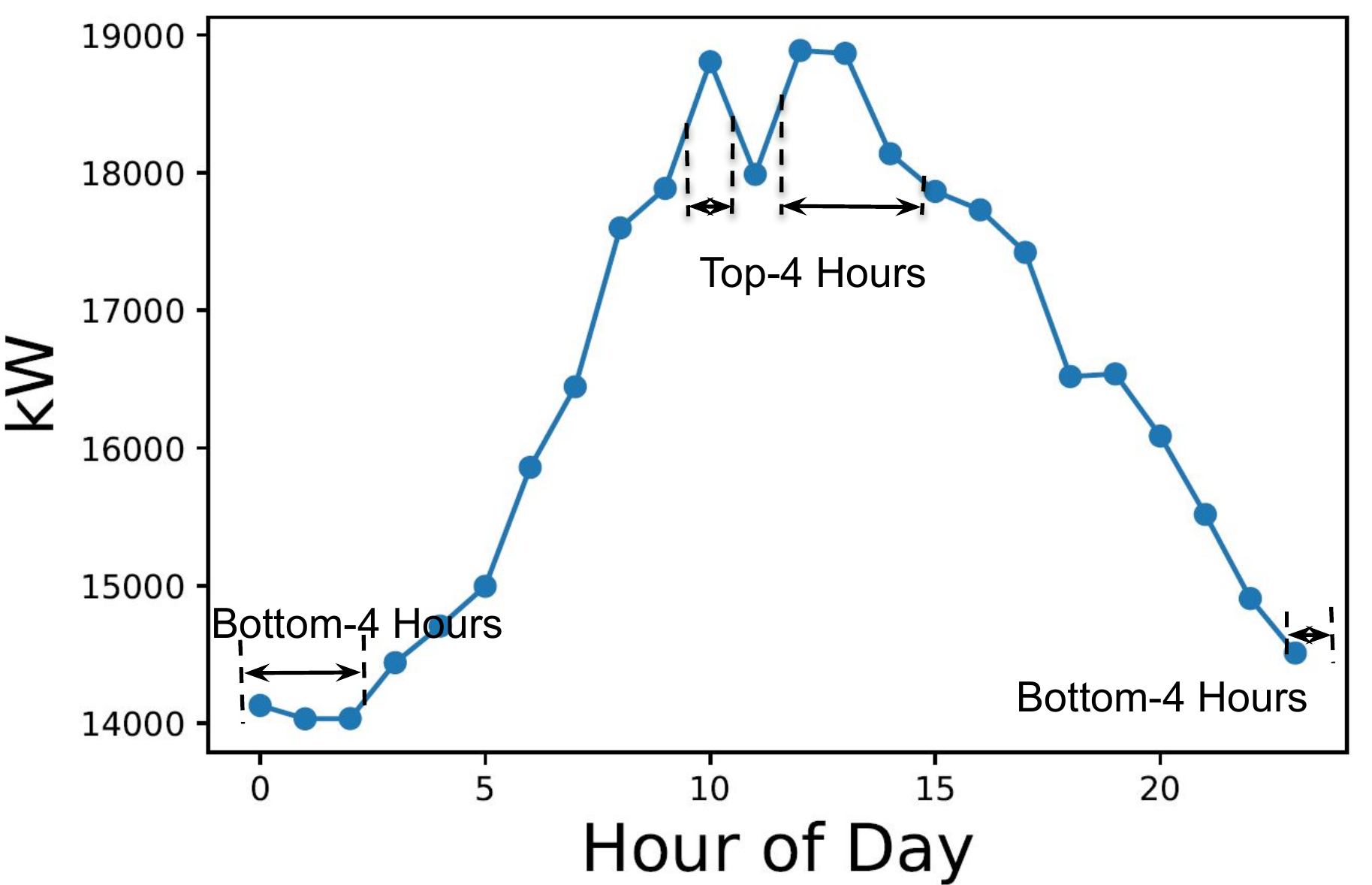}\\
    (a) & (b)
    \end{tabular}
    }
    \vspace{-0.3cm}
    \caption{ Sample peak forecasts for (a)uni-modal demand peak and (b) bi-modal demand peak.}
    \label{fig:model_compare}
    \vspace{-0.4cm}
\end{figure}

\subsection{Peak Predictions}
Next, we show the sample output of our LSTM model for forecasting top and bottom-k hours. 
Figure \ref{fig:model_compare} show the campus micro-grid demand on two different days and shows the labeled top-k and bottom-k hours in each day. 
Note that some days can be unimodal where the top/bottom $k$ hours occur contiguously, while other days can be bimodal where these hours are non-contiguous. Our model can handle both scenarios, as shown.



\subsection{Peak Forecast Accuracy}
Next, we compare the peak forecasting accuracy of various models. 
Tables ~\ref{tab:topk} and ~\ref{tab:bottomk} show the accuracy of various approaches for predicting the top-k and bottom-k hours of each day, respectively, for various values of $k$. We evaluate model accuracy as the percentage of the correct number of peak hours captured by each model.


\begin{table}[t]
    \centering
   \begin{tabular}{lccccc} \toprule
        Model &  1 & 2 & 3 & 4 & 5 \\ \midrule
        LSTM-2 Layer & 16\% &    26\%  &    74\%  &    79\%  &    84\% \\
        \textbf{LSTM-4 Layer} & \textbf{47\%}  & \textbf{74\%} & \textbf{89\%} &    \textbf{95\%} & \textbf{100\% } \\
        Linear Regression & 42\%  & 68\%  & 84\%  & 89\%  & 100\% \\
        ANN (SOTA) &  26\%  & 42\%  & 53\%  & 63\%  & 79\%\\
        Custom ARIMA \cite{kim2019short} &  42\%  & 63\%  & 74\%  & 84\%  & 95\% \\ \bottomrule
    \end{tabular}
    \caption{Top-k peak prediction accuracy, k range 1 to 5}
    \label{tab:topk}
    \vspace{-0.5cm}
\end{table}

\begin{table}[t]
    \centering
   \begin{tabular}{lccccc} \toprule
        Model &  1 & 2 & 3 & 4 & 5 \\ \midrule
        LSTM-2 Layer & 42\% & 50\% & 58\% & 75\% & 92\% \\
        \textbf{LSTM-4 Layer} & \textbf{42\%} & \textbf{50\%} & \textbf{67\%} & \textbf{83\%} &\textbf{92\%} \\
        Linear Regression & 33\% & 42\% & 50\% & 58\% & 67\% \\
        ANN (SOTA) & 25\% & 58\% & 67\% & 75\% & 83\% \\
        Custom ARIMA \cite{kim2019short} & 33\% & 42\% & 50\% & 83\% & 92\% \\ \bottomrule
    \end{tabular}
    \caption{ Bottom-k Peak prediction accuracy, k range 1 to 5}
    \label{tab:bottomk}
    \vspace{-0.75cm}
\end{table}

First, we observe that the accuracy is lower for small values of $k$ since the chances of making mistakes is higher for small $k$ (e.g., if the top two hours are close to one another, for $k=1$, a model may choose the wrong hour due to prediction error). The accuracy of all approaches increases with a higher $k$ since each approach only needs to find all hours in the top $k$ regardless of the order.  The table shows that the 4 layer LSTM approach outperforms all other approaches. For $k\geq4$, it yields an accuracy of 95\% and 100\%, respectively. For $k=1$, its accuracy is only 47\% but it is still greater than all other approaches. Overall, it has 11-32\% better accuracy than the state of the art ANN and Custom ARIMA approaches.  Interestingly, linear regression performs quite well despite its simplicity and is able to outperform ANN and custom ARIMA for top-k predictions.

Table \ref{tab:bottomk} shows bottom-k prediction accuracy. Note that top-k forecasts are more critical than bottom-k predictions since an error in top-forecasts directly impacts incurred the demand charges, while error in bottom-k forecast implies that the battery may charge in different hours than the true bottom and we can tolerate more errors in bottom $k$ predictions. Again, 4 layer LSTM approach outperforms all other approaches. However, its accuracy is slightly lower than when making top-k forecasts. The accuracy of all methods increases for higher values of $k$, like before.  However, the two states of the art methods and the linear regression have much lower accuracy than our LSTM models.

\subsection{Prototype Implementation and Efficiency}
We have implemented a full prototype of our peak forecasting model.
One of the goals of our work is to develop compact, efficient models that can be deployed and executed on embedded processors that are common in the battery control system---with the goal of using the models to drive autonomous operation of batteries for peak shaving.  Our 4 layer model has a memory footprint of  ~300KB, while the 2 layer model has footprint ~150KB, which allows them to fit into low-end devices with small amounts of RAM. We ran both models on a Raspberry PI 3 device running 32-bit Linux and measured the execution latency. Both are able to predict the top and bottom 5 peaks of an entire day in less than 1.8 seconds. 
We have released our model as an open-source library on GitHub: \url{http://github.com/umassos/peak-prediction} 

\section{Peak Shaving Case Study}
\label{sec:casestudy}

\begin{figure}[t]
\centering
\includegraphics[width=2in]{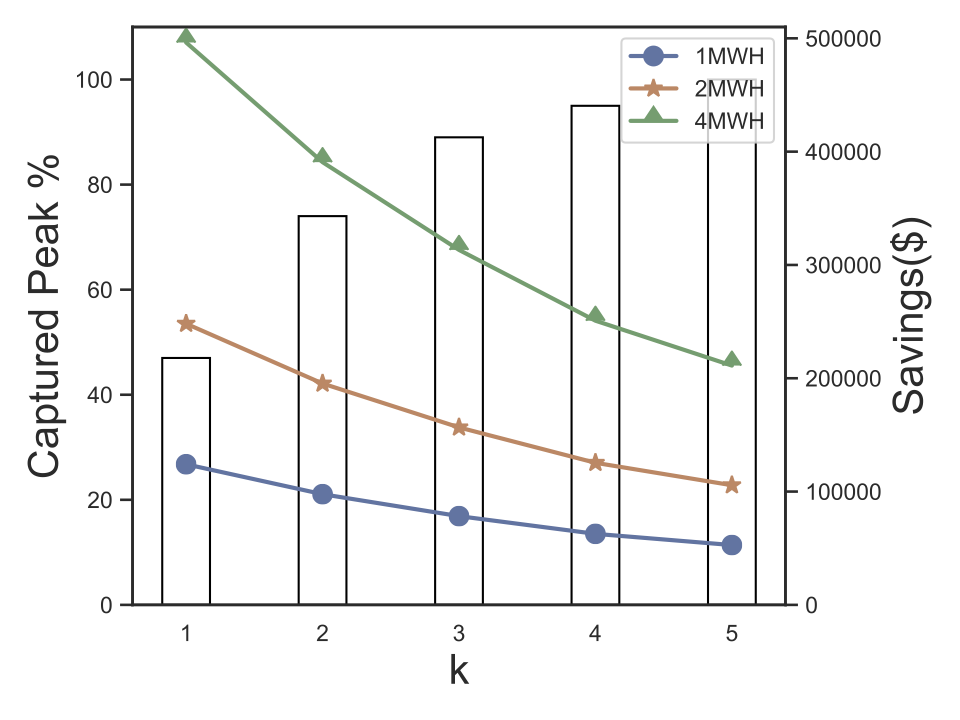}
\vspace{-0.4cm}
\caption{Battery payback and savings for varying battery sizes}
\label{fig:payback}
\vspace{-0.57cm}
\end{figure}

Finally, we present a case study to evaluate the efficacy of our peak forecasting for peak shaving. To do so, we assume that the model peak predictions are used to directly drive the battery charging and discharging.  
To evaluate the overall cost efficacy of our approach, we consider various battery sizes (1MWhr, 2MWhr, 4MWhr). 
For each battery size, we compute the savings over a period of k hours per day, where k $\in$ \{1,2,3,4,5\} represents the number of hours of battery discharge operation per day. Since the model captures all peaks at k=5 we compute the savings till k=5.  The per-hour battery discharge (BD) for a specific value of $k$ is assumed to be $\frac{Battery\_Capacity}{k}$.

Our campus has recently installed a 4MWhr battery for peak shaving. The local utility company imposes a \$22/kW demand charge for usage during peak hours. The cost of a 4MWhr battery installed is approximately \$800,000, with a unit cost of \$200/kWhr. Figure \ref{fig:payback} shows the computed savings in demand charges for all 3 battery sizes. We find that k=1 gives us the best savings and payback irrespective of the battery size. The model accuracy in predicting the peak increases with increasing values of k. So, model accuracy is higher for k=2 than k=1 but as k increases the battery discharge per hour reduces. This drop in battery discharge offsets the cost savings substantially as k becomes greater than 2 resulting in a drop in the savings. The figure shows annual savings of \$496,320 for a 4 MWhr battery for $k=1$ and annual savings of nearly \$200,000 for $k=5$.
Overall the figure shows the efficacy of our peak forecasting approach for extracting real-world savings by shaving peak demand.  Based on this case study and the above experiments, we are deploying of our model for the day-to-day control of the campus battery for peak shaving.





%% file: conclusions.tex
\section{Conclusions}
\label{sec:conclusions}

In this paper, we presented a peak forecasting model for battery-based peak shaving.  Our deep learning-based model predicts the top and bottom-k hours of each day, which can then be used to control a battery for peak shaving.  We showed that our approach outperforms the state of the art load forecasting techniques adapted for peak predictions by 11-32\%. When used for battery-based peak shaving, our model yields annual savings of \$496,320 for a 4 MWhr battery for a campus micro-grid.

\textbf{Acknowledgements:} This research is supported by NSF grant 1645952, DOE grants DE-EE0007708, DE-EE0008277, and MA Department of Energy Resources and MA Clean Energy Center.